\documentclass[showkeywords,preprint,pre,onecolumn]{revtex4}
\usepackage{hyperref}
\usepackage{graphicx}
\usepackage{amsmath}
\usepackage{amsmath}
\usepackage{amsfonts}
\usepackage{amssymb}
\usepackage{latexsym}
\usepackage{ifpdf}

\pdfoutput=1
\begin{document}

\title{Building Cooperative Networks}

\begin{abstract}

We study the cooperation problem in the framework of evolutionary game theory using the prisoner's dilemma as metaphor of the problem. Considering the growing process of the system and individuals with imitation capacity, we show conditions that allow to form highly cooperative networks of any size and topology. Introducing general considerations of real systems, we reduce the required conditions for cooperation to evolve approaching the benefit-cost ratio $r$ to the theoretical minimum $r=1$, when the mean connectivity of the individuals is increased. Through the paper, we distinguish different mechanisms that allow the system to maintain high levels of cooperation when the system grows by incorporation of defectors. These mechanisms require heterogeneity among individuals for cooperation to evolve. However, the required conditions and heterogeneities are drastically reduced as compared to those required for static networks.
 
\bigskip 
\bigskip 


\end{abstract}

\author{Ignacio Gomez Portillo}
\affiliation{Grup de F\'isica Estad\'istica, Departament de F\'isica, Universitat Aut\'onoma de Barcelona, 08193 Barcelona, Spain.}
\maketitle

\section{Introduction}

Cooperation  \cite{G1,G2,G3,G4} is present in every form of life, some examples are the cells working together in order to form multicellular organisms and these working together to form societies such as anthills and the human society. It is noteworthy that cooperating biological systems make up new organisms that in turn can cooperate to form a new one with higher complexity. This shows that cooperation is a mechanism through which nature increases complexity of life. In this way, if there are organisms made up by other cooperative organisms, what would happen when for some reason these stop cooperating? What would happen to us if our cells stop cooperating? We easily note that the lack of cooperation means the death of the most complex organism. Therefore, the emergence of cooperation produces a new biological organism that dies when cooperation disappears. 

\bigskip

Understanding how cooperation emerges and persist in the framework of Darwinian Theory is a big challenge that has received attention for decades \cite{G5,G6,G7,G8,G9,G10,G11,G12,G13,G14} but still has problems to be solved. Evolutionary game theory \cite{G15,G16} has proved to be an appropriate theoretical framework to formally represent the problem. In this theory, the interactions between two individuals are represented by a game in which each player adopts a strategy. The system evolves replicating more often the strategies of successful individuals in the game. In particular, the Prisoner's Dilemma has been the most widely studied game as metaphor of the cooperation problem. In this game, each player adopts one of the two possible strategies, cooperation ($C$) or defection ($D$). When in an interaction both individuals cooperate each receive a Payoff $R$ and other one $P$ under mutual defection. If one cooperates and the other defects, the former receives $S$ and the second $T$. The game is completely defined when $T>R>P>S$. Under these conditions and in a one-shot game, it is better to defect no matter the strategy adopted by the opponents. Thus, in fully connected systems defection always has the highest reproduction rate. Therefore, the system evolves decreasing the fraction of cooperators to extinction. This highlights the difficulty of the cooperation problem when it is framed within the Darwinian Theory. 

\bigskip 

Until the early nineties the problem was only considered in fully connected systems. So, to overcome the problem it is necessary to introduce features to individuals that increase the strategic complexity of the game. In $1981$, Axelrod and Hamilton \cite{G6} showed that cooperation based on reciprocity evolves when interacting individuals know that the probability of new encounters is high enough. In addition, other possibilities have been considered, as individuals with reputation \cite{G10} and punishment for defectors \cite{G14}. In a pioneering work \cite{N1}, Nowak and May studied the problem over regular lattices, where all individuals have an equal number of neighbors $k<N-1$, with $N$ the size of the system. In this case, each individual is limited to interact only with its neighbors rather than with the whole system. In this way and without increasing the strategic complexity, they showed that there are conditions under which cooperation is not extinguished. This shows us that locality of individuals is fundamental to the evolution of cooperation as well as the need to introduce new topological features of real systems. However, the lack of knowledge of these structures delayed until the beginning of the third millennium the publication of new works following this way \cite{N2,N3,N4,N5,N6,N7,N8,N9,N10,N11,N12,N13,N14,N15,N16}. In $2005$, Santos \emph{et. al}. \cite{N3,N4} showed the importance that the degree heterogeneity may have \cite{N13,N14,N15} for cooperation. In particular, the scale-free degree distributions are those that most favor cooperation in static systems and when the effective Payoff of individuals is the accumulated one from interaction with their neighbors. These systems are characterized by a distribution of the number of links k, governed by a power law $P(k)\sim k^{-\gamma }$, with $2<\gamma <3$. This is an outstanding result because scale-free behavior is common in biological and technological systems \cite{N20}. Also other typical structures of real systems such as high clustering coefficient \cite{N5,N6} and degree-degree correlations \cite{N10} have been considered.

\bigskip 

These works \cite{N1,N2,N3,N4,N5,N6,N7,N8,N9,N10,N11,N12,N13,N14,N15} assume static networks large enough to ensure that the topological features under study are well developed. Initially each individual adopts cooperation o defection with equal probability, in order to not favor any strategy. Then the system evolves following an evolutionary game until it reaches the steady state, characterized by a stable fraction of cooperators. Although this form of modeling the cooperation problem introduces significant features of real systems it has some limitations. In particular, cooperative biological systems do not emerge spontaneously fully developed but they are formed by a growing process from one or few individuals. Furthermore, it is remarkable that in all stages the system has levels of cooperation higher than half of the population. These features have important consequences for the conditions required for cooperation to evolve when individual have imitation capacity \cite{N17}. Therefore, it is necessary to carry out extensive studies considering the growing process as an essential ingredient for cooperation. Furthermore, another feature of any biological systems is the possibility of mutation, which in the context of cooperation means a spontaneous change of strategy. Therefore, taking into account the growing process and the possibility of mutations, we formulated the cooperation problem as follows:

\bigskip

- What conditions should the system have to grow as a highly cooperative system?

- What are the system properties that allow to overcome the emergence of mutant defectors in a cooperative system?

\bigskip

In this paper we explore the cooperation problem considering the growing process of the system. It is noteworthy that this process has also been considered \cite{C5,C6} as a mechanism to introduce co-evolutionary dynamics \cite{C1,C2,C3,C4,C5,C6}, where the strategies of individuals and topological features are correlated. We consider the topological structure and the evolution of strategies decoupled. We focus our attention in the first question assuming that the probability of mutation is low enough to be neglected, this is a routinely simplification in literature \cite{N1,N2,N3,N4,N5,N6,N7,N8,N9,N10,N11,N12,N13,N14,N15,N16,N17}. As it was previously mentioned, the result of interactions of each individual defines its Payoff which increments its fitness. Therefore, natural selection is introduced through payoff since individuals with a higher one reproduce more. The reproduction can be either genetic or cultural. When it is genetic, the most successful individuals have the most offspring. When reproduction is cultural the strategies of successful individuals are more likely to be imitated by their neighbours. In this work, unless otherwise indicated, we implement the evolutionary game theory for cultural reproduction. Therefore, we are considering that individuals have imitation capacity.

\section{The model}

The system is represented by a network of interconnected nodes. Each node is an individual who interacts with those who are connected with it. Unless otherwise stated, we considered all nodes equal and all connections undirected and with equal weight. Each interaction is modeled by a round of the prisoner's dilemma game. We choose $T=b$, $R=1$, $P=0$ and $S=-c$ to reduce the analysis to a single parameter defined by the benefit-cost ratio $r=b/c$. Strategies evolve by the finite population analog of replicator dynamics \cite{N3,N4} for the prisoner's dilemma. To compute the Payoff $P_{i}$ of individual $i$ we perform the following. If $i$ is a cooperator, after interacting with their neighbors it obtains a Payoff $P_{i}=k_{i}^{c}b-k_{i}c$, where $k_i$ is the degree of $i$ and $k_{i}^{c}$ the number of cooperative neighbors. When $i$ is a defector linked to $k_{i}^{c}$ cooperators it gets a Payoff $P_{i}=k_{i}^{c}b$. After an interaction for each connection of the system, all strategies are updated simultaneously (synchronous update) as follows: Each individual $i$ randomly chooses a neighbor $j$ and compares $P_{i}$ with $P_{j}$. If $P_{i}\geq P_{j}$ the player $i$ keeps its strategy. On the other hand, if $P_{i}<P_{j}$ the player $i$ adopts the strategy of its neighbor $j$ with probability $P_{j\rightarrow i}=(P_{j}-P_{i})/k_{m}(b+c)$, where $k_m=max\{k_i;k_j\}$. The denominator ensures a proper normalization of the transition probability. A complete update of the system is called a generation. Furthermore, the results reported throughout the paper are qualitatively equivalent when strategies are updated in an asynchronous way or when \cite{N17} the update rule is replaced by the Fermi rule \cite{N17,C5,C6} for all non-zero natural selection intensity. The latter is important because shows that conclusions are robust to error desicions of the individuals since the Fermi rule allows that individual $i$ takes strategy of $j$ even when $P_{i} \ge P_{j}$.

\bigskip

To take into account the growing process and the strategies update simultaneously it is necessary to define a temporal scale. We accomplished this through the system size assuming that $\frac{dN(t)}{dt}=aN(t)$, where $a$ is the growth rate. When the elimination of nodes is not considered, the growth rate is equal to the frequency with which individuals are incorporated to the system, this is considered constant over time. Therefore, the population grows exponentially as $N(t)=N(t_0)e^{a(t-t_0)}$. Between two generations the system grows a time $\Delta t$. So if a strategy update is performed when the system has a population $N(t_0)$, then the next one is performed when the system has a population $N(t_0+\Delta t)=N(t_0)e^{a\Delta t}=N(t_0)(1+n)$. Consequently, the strategies are updated each time the system grows a fraction $n$, it is important to note that $a\Delta t\simeq n$ when $a\Delta t<<1$. Unless indicated otherwise, results are obtained for $n=0,001$. It has been checked that the results do not change significantly for $n\in \lbrack 0,01;0]$. In particular, as it has been shown in \cite{N17}, there is an enhancement of the cooperation level into the system when $n$ decrease. At some $n_{c}>0,01$ the number of defectors who join the system before the next generation increase the required conditions for cooperation to prevail. Moreover, from some $n$ the number of defectors that become cooperators in a generation is less than the number of defectors introduced to the system in the same time. Therefore, the system is not able to sustain cooperation under any set of conditions. Finally, the model can be extended to considered other kinds of growth as for instance the logistic one, where the system initially grows exponentially in time but then it slows down until the system acquires a maximum size called carrying capacity of the system. Under this kind of behavior, it is expected that the frequency with which new individuals are incorporated to the system decreses in time and, therefore, the growing proportion $n$ before a strategy update also decrease. Therefore, the qualitative conclusions reached through the paper do not change when it is performed this kind of considation.

\bigskip

Actual cooperative systems grow from one or a few individuals by incorporating new ones through two different mechanisms. On one hand, the new individuals may come from the reproduction of individuals of the system and, therefore, related genetically. On the other hand, the system grows by the incorporation of independent individuals who could not be related genetically to existing individuals. However, unless otherwise specified, the proportion of new individuals coming from each of the two sources does not change the results. Throughout the paper we use different models for the growth of the network; they are justified and detailed when necessary. Finally, we must specify how individuals adopt the strategy in the first interaction with its neighbors. The Darwinian Theory restricts how strategies evolve but gives complete freedom in the initial strategy of each individual. We define $P_{c}$ as the probability that an individual is initially cooperator. In the literature \cite{N1,N2,N3,N4,N5,N6,N7,N8,N9,N10,N11,N12,N13,N14,N15}, it is often taken $P_{c}=1/2$ in order to not favor any strategy. For our purposes we can relax the condition to just $P_{c}\neq 0$ in order to introduce cooperation into the system.

\bigskip

Throughout the paper we compare the results for the growing systems with those found in static systems \cite{N3,N4,N5}. This allow us to show the importance of the growing process of the system for cooperation. The results for static systems were obtained by considering a population $N=10^{4}$. Initially we randomly distributed cooperators and defectors. Then the system evolves following the evolutionary dynamics specified in the first paragraph of this section. The reported results have been obtained when the evolutionary dynamic acquires a stable state. This has been defined as follows when the system is static and when it grows:

\bigskip

Static systems: The stationary regime is characterized by the average fraction of cooperators $\bar{c}$ the system. The value of $\bar{c}$ is computed as follows. The system evolves a transient of $\Delta g=10^{4}$ generations. Then we study the evolution of the number of cooperators $c(g)$ of each generation $g$ during time windows of $\Delta g=10^{3}$ generations by computing the average value and fluctuations of $c(g)$. When the fluctuations are less than or equal to $1/\sqrt{N}$ we stop the simulation and calculate $\bar{c}$ of the realization as $\bar{c}=\Sigma_{g}c(g)/(\Delta gN)$. The reported values ​​have been obtained by averaging $100$ realizations corresponding to $10$ different initial conditions of the strategies for $10$ different networks.

\bigskip

Growing systems: The system evolves following the model up to $N=10^{4}$. Then, we stop the growth and calculate $\bar{c}$ using the same stability criterion used for static systems. We adopt this procedure for two reasons. On the one hand, it allows us to make a correct comparison of results with those obtained in static systems because both values are ​​obtained by the same procedure. On the other hand, growing systems have a phase transition from non-cooperative to cooperative behaviors for a $r$ critical $r_{c}$. As $r\rightarrow r_{c}$ the levels of cooperation within the system are modified each time more slowly as the system grows. Therefore, to properly characterize the transition as the system grows, it is necessary to reach increasingly larger systems making the task difficult. It has been checked that the conclusions reached are not altered if $\bar{c}$ is calculated by applying a criterion of stability as the system grows. The results reported are obtained by averaging over $40$ realizations.

\bigskip 

\section{Results}

\subsection{Model analysis}

To perform a clear analysis from the literature \cite{N1,N2,N3,N4,N5,N6,N7,N8,N9,N10,N11,N12,N13,N14,N15,N16} we study the model as follows. We assume that there are conditions under which the system reaches a state of $N_i$ cooperators. We take $N_i$ large enough to ensure that the macroscopic topological structures under study, such as the degree distribution, are well developed. From this structure, we seek the conditions required to maintain a stable and high level of cooperation when the system grows by the incorporation of defectors. This simulates the worst condition that the cooperative system must resist. Using the conditions found we look for the minimum $N_i=N_c$ of cooperators that ensure the stability of the cooperation level within the system. We carry out this analysis for two different growth mechanisms yielding realism to the network generation model. This procedure reduces the minimal conditions to keep cooperation stable. Then, we study the model when the cost to maintain a link is considered. For this purpose we make changes to the way individuals accumulate Payoffs. Finally, we focus our attention on the first individuals of the system showing conditions that allow us to form a cooperative structure of $N_c$ cooperators as well as to form cooperative systems of any size and topology.

\subsection{The model in Barabasi-Albert scale-free networks}

As we previously mentioned in the introduction, the degree heterogeneity plays a key role for cooperation when the problem is modeled by the evolutionary dynamic discussed in Section II, but applied in static networks. In this case, the scale-free degree distributions, very abundant in nature, are the most suitable ones for cooperation to evolve. Taking this as motivation and to make an appropiate comparison of results with those obtained in previous works \cite{N3,N4,N5}, we began studying the model considering a growing process of the system given by the Barab\'{a}si-Albert Model (BAM) \cite{N18}. In this model, the system grows from $N_{0}$ fully connected individuals through the incorporation of new individuals with $L\geq N_{o}$ new links, we consider $L=N_{0}$. Each new connection is performed by preferential attachment \cite{N18} with nodes of the system. These incorporate a new connection with a probability proportional to its degree and avoiding double connections with the new individual. This procedure produces scale-free networks with exponent $\gamma \simeq 2,9$ in the thermodynamic limit $N \rightarrow \infty$.

 \bigskip

\begin{figure}[!hbt] \centering
\includegraphics[angle=0,scale=1]{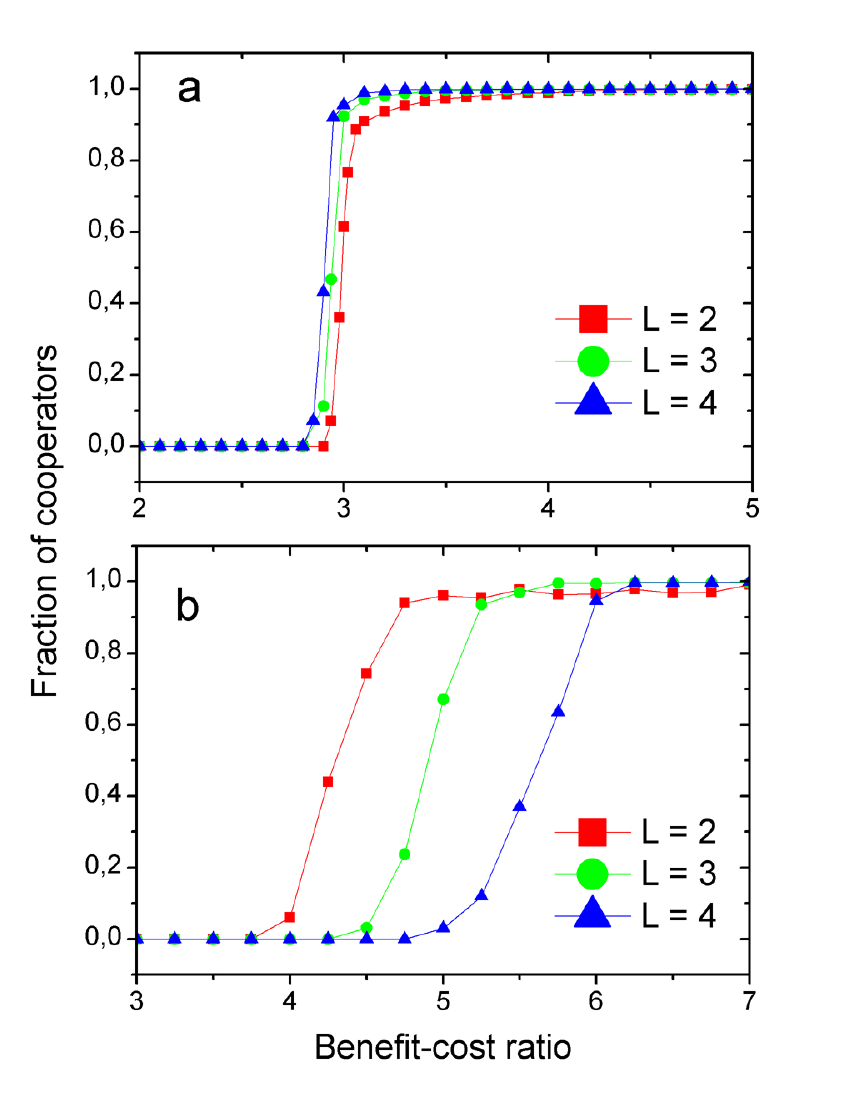}
\caption{The model in Barabasi-Albert scale-free networks. Fraction of cooperators $\bar c$ as a function of the benefit-cost ratio $r$ for differents values of $L$. a) Growing systems. b) Static systems.}
\end{figure}

\bigskip 

In Figure $1a$ we show the fraction of cooperators as a function of the benefit-cost ratio $r$ for different values ​​of $L$. As it shows, the system exhibits a phase transition from a non-cooperative to a cooperative state when $r$ exceeds the critical threshold $r_{c}\simeq 2,9$. We note a slight decrease of $r_{c}$ when $L$ increases. Nonetheless, this behavior becomes less apparent when $L$ increases. We have verified this behavior to $L=32$. Therefore, considering that the mean degree $\bar{k}\simeq 2L$ when $N>>L$, we may consider $r_{c}$ independent of $\bar{k}$ when the system grows by BAM. This can be understood by approaching $\bar{c}$ by $1$ when $r>r_{c}$. Under this condition the new defectors receive a Payoff proportional to $L$ given that all its neighbors are cooperators. Furthermore, we can assume that the new defector exploits cooperators with an average Payoff proportional to $\bar{k}$. Therefore, given that the relation $\bar{k}/L\simeq 2$ is independent of $L$, it is expected $r_{c}$ to be independent of $\bar{k}$. Moreover, the striking similarity between the exponent of the degree distribution and $r_{c}$ motivates analytical studies in order to explore a possible relationship. Finally, for $r>r_c$ there is an increment of the fraction of cooperators when $L$ increases, causing a slight increase in $r_{c}$ for small $L$.

\bigskip 

In Figure $1b$ we show analogous results obtained over static networks. Comparing these with those shown in Figure $1a$ we note a significant decrease of $r_c$ when it is considered the growing process of the system. The origin of this difference is determined by two fundamental characteristics caused by the growing process. On the one hand, the system always has high levels of cooperation instead of initially half of the population considered in static systems. On the other hand, cooperators inhabit the most connected parts of the system rather than being initially randomly distributed. This reduces the exploitation capacity of defectors to a minimum because it is proportional to its degree $k$. Also, when we considered static systems, there is an increase of $r_{c}$ to growing $L$. We have corroborated this behavior for values ​​of $L$ higher (up to $L=32$) than those shown. Thus, the benefits from the system growth increase with $\bar{k}$. Therefore, taking into account Figure $1a$ and $1b$ we shown that the growing process is fundamental to the evolution of cooperation \cite{N17}. It is important to note as it has been shown previously \cite{N5,N8,N16}, that the conditions required for cooperation to evolve increase with increasing $\bar{k}$ in static systems. However, it is curious to note that an opposite behavior has been reported in \cite{N3},  making furhter checks necessary.

\bigskip 

\begin{figure}[!hbt] \centering
\includegraphics[angle=0,scale=1]{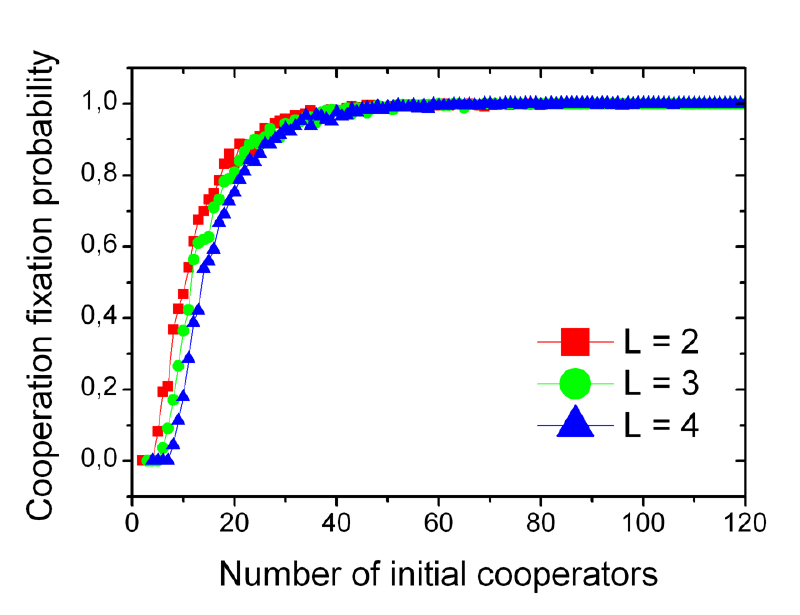}
\caption{The model in Barabasi-Albert scale-free networks. Cooperation fixation probability as a function of the number of initial cooperators for differents values of $L$. We have considered $r=3,5$ and $M=500$.}
\end{figure}

In Figure $2$ we show the cooperation fixation probability $P_{f}(N_{i})$ for different values ​​of $L$, where $P_{f}$ is defined as the probability that a system of $N_{i}$ cooperators continues being cooperative when it grows by the incorporation of defectors. To obtain $P_{f}$ we select a value $r>r_{c}$ and we perform $M$ simulations starting from a system of $N_{i}$ cooperators. Then, we compute the number $M_{c}$ of systems that reach $N=10^{4}$ with a fraction of cooperators $\bar{c}>1/2$. Finally, we compute $P_{f}$ for each $N_{i}$ as $M_{c}/M$. As it can be seen in all curves, $P_{f}$ grows steeply and reaches the value $1$ for a size $N=Nc\sim 100$. From this structure, called \emph{Cooperative Seed} (CS) \cite{N17}, the cooperation is stable despite the system grows by incorporating defectors. When the system is small, the degree distribution may not be well developed, thus reducing the degree heterogeneity between the new individuals and the rest of the system, necessary \cite{N17} to resist the defection invasion introduced by the new individuals. Therefore, the cooperation can be extinguished because of finite-size effects when $N<N_{c}$. We have also checked that $N_{c}$ decreases as $r$ increases because less degree heterogeneity causes an increase in $r_{c}$, as shown in both growing \cite{N17} and static \cite{N3,N4,N5} systems. One could expect that $N_{c}\rightarrow \infty $ when $r\rightarrow r_{c}$, but it is found that the results shown in Figure $1a$ do not change significantly when $N_i> 200$. Therefore, it is a good approximation to assume $r_{c}\neq r_{c}(N)$. In static systems the conditions for the evolution of cooperation are valid \cite{N3} from $N_c \sim 100$. This is important because it assures us to be making a correct comparison of the results. The \emph{Cooperative Seed} can be justified considering the probability $P_{N_{c}}=(P_{c})^{N_{c}}\neq 0$ to have the first $N_{c}$ cooperators. However, unless $P_{c}\rightarrow 1$, the cooperative seed size leads to a very small probability $P_{N_{c}}$ that makes difficult the formation of cooperative systems. Therefore, we have to introduce new considerations to reduce the \emph{Cooperative Seed}.

\subsection{Model A and its consequences for cooperation}

In this subsection we modify the growth mechanism BAM making considerations ​​on the formation of links and the incorporation of individuals. For this we have the caution of not losing the scale-free behavior introduced by BAM. In this, each new individual performs $L$ simultaneous connections to the system. However, in real systems the individuals do not make several connections simultaneously. Instead, each connection is incorporated at different times. This feature is important because it gives the possibility of strategy updates when individuals have less than $L$ connections and therefore reduces the exploitation capacity of those who are defectors. We introduce this by considering the formation of new links within the system at the same time that new individuals join it. Below we introduce the \emph{Model A} (MA) for network generation.

\bigskip 

\begin{figure}[!hbt] \centering
\includegraphics[angle=0,scale=0.3]{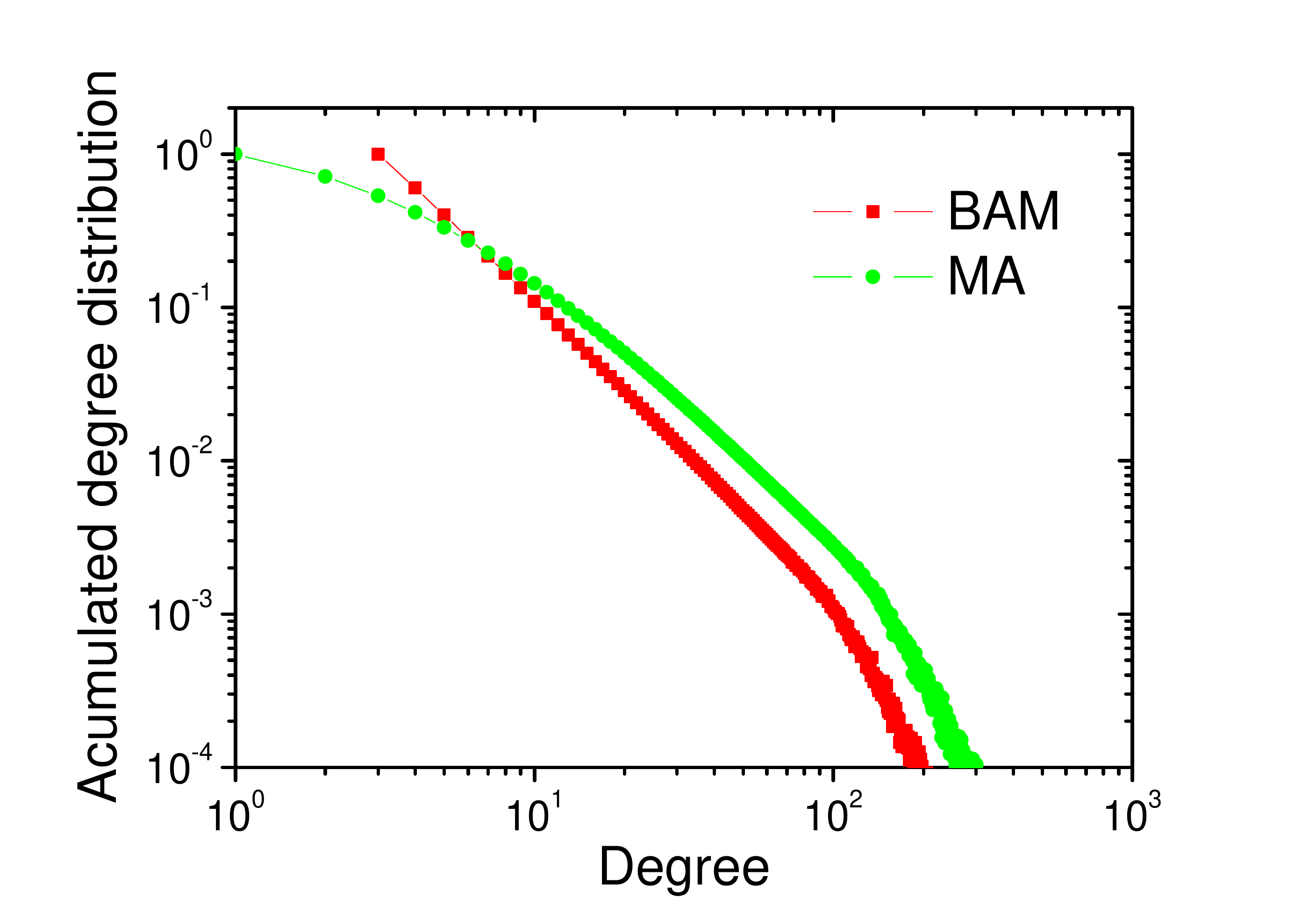}
\caption{The model A. Accumulated degree distribution $P_{a}(k)$ for $L=3$ and $N=10^4$. The reported values have been obtained by averaging over $100$ different networks. }
\end{figure}

\bigskip 

Just as in BAM, the system grows by incorporating new individuals from a fully connected network of $N_0$ individuals. In addition, each new individual incorporates to the system $L=N_0$ links. But instead of all new links belong to the new individual, we perform a splitting of the $L$ connections. On the one hand, the new individual performs its first link (FL) with the system through preferential attachment. On the other hand, the remaining $L-1$ links (RL) are performed by choosing a random individual who makes a connection to any other individual of the network chosen by preferential attachment. To see how the splitting of the $L$ connections affects the degree distribution $P(k)$, we placed in Figure $3$ the accumulated degree distribution $P_{a}(k)=\int \limits_{k}^{\infty }P(k)dk$ emerged from MA for $L=3$. Also, we show the analogous result obtained by BAM. As shown in it, both models have in their tails equivalent behavior, decaying as a power law with exponent $\gamma =2,9$ followed by a faster decay after some degree $k$. The latter effect is caused by a finite size effect of the system. However, in contrast to BAM, Model A allows nodes with degree $k<L$ but preserving the mean degree $\bar{k}\simeq 2L$. This generates a decrease in the number of links among individuals with $k < \bar{k}$ which is compensated by an increase in the connectivity of individuals with degree $k>\bar{k}$. This fact is reflected through the bending of the degree distribution for small $k$ and a shift of the scale-free behavior to the right in the degree distribution of MA. It is noteworthy that the bending for small $k$ is a typical feature in real systems \cite{N20} and, therefore, through MA we have introduced greater realism the model. 

\bigskip 

The simple modification introduced through MA allows some reinterpretations in the way that the RL are generated. In BAM, one of the ends of the RL is always connected to the new individual. But now both link ends can be attached to any individual of the system. This symmetry allows us interpret that the RL are performed as follows. First, we randomly choose an individual $i$ of the system. This performs one of the $L-1$ RL with probability $k_{i}/\sum_{j}k_{j}$ with another randomly chosen individual of the system. We repeat the procedure until all ​​the $L-1$ links have been made. Thus the connection instead of ending on a preferential way, is born preferably from well-connected individuals. This attributes to the connections between individuals the capacity of reproduction. This is of interest because it suggests a possible explanation for the origin of the preferential attachment required for scale-free behavior of real systems. Also, this could be an important mechanism for the evolution of cooperation taking into account that cooperative links are the most successful ones and therefore with higher reproduction rate. This kind of interpretation may also be justified from a biological point of view considering that a set of cooperating individuals form a new individual and, therefore, with the potential to reproduce. For a complete analysis of Model A as well as other alternatives on the same line we refer to \cite{N19}.

\bigskip 

\begin{figure}[!hbt] \centering
\includegraphics[angle=0,scale=1]{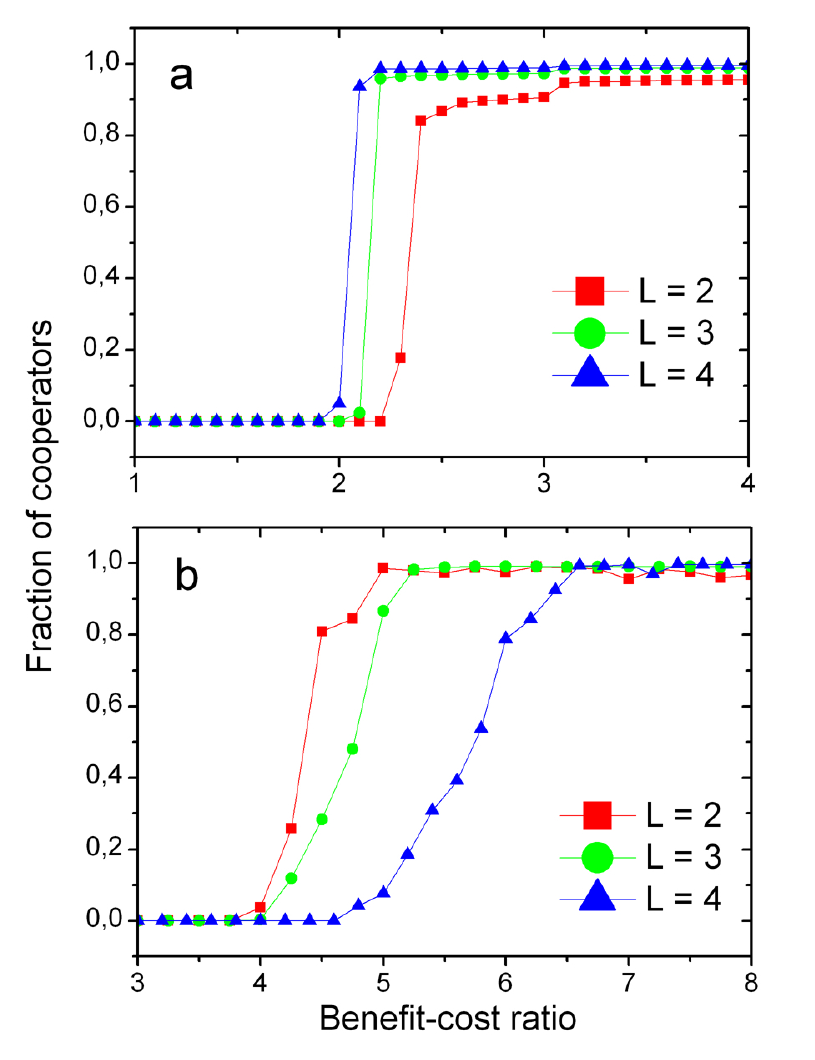}
\caption{The model over networks generated by MA. Fraction of cooperators $\bar c$ as a function of the benefit-cost ratio $r$ for differents values of $L$. a) Growing systems. b) Static systems.}
\end{figure}

In Figure 4a we show the fraction of cooperators as a function of $r$ for different values ​​of $L$ when the system is generated by MA. As expected, the system shows a phase transition from a non-cooperative to a cooperative behavior for some $r$ critical $r_{c1}$. It is remarkable that $r_{c1}$ is significantly lower than the $r_c$ obtained by BAM, showing the importance to the cooperation of the splitting performed to the $L$ links. As we can observe, the value of $r_{c1}$ decreases when $\bar{k}$ increases. We have checked that this behavior still makes $r_{c1}$ tend to the theoretical minimum $1$ (for $L=32$, $r_{c}\simeq 1.3$). The reduction of $r_{c1}$ arises because the increase of $\bar{k}$ enhances the average differences between the connectivity of new individuals and the rest of the system. Also, surprisingly, the system exhibits a new phase transition that slightly increases the level of cooperation with critical threshold $r_{c2}$. This shows that the system has different mechanisms to resist the invasion of defectors and which are activated at different $r_{c}$. It is important to note that $r_{c2}$ matches with $r_{c}$ obtained when the system grows by BAM. On the one hand, this similarity shows that $r_{c2}$ is determined by the topological features shared by BAM and MA, namely the tails of the degree distributions. On the other hand, the topological difference between the two network models, namely the curvature of the distribution for low degree, produces the appearance of a new mechanism for cooperation to evolve. However, the underlying processes by which the system resists the invasion of defectors are still unclear and of great interest. Therefore, it is necessary to perform deeper analyses to fully understand these phenomena. It is important to note that it has been checked that this new transition is not a consequence of the stability criterion adopted.

\bigskip 

In Figure $4b$ we place $\bar{c}(r)$ obtained when we consider static systems generated by MA for different values of $L$. Again, the system exhibits a phase transition from a non-cooperative to a cooperative behavior for some $r_{c}$. It is remarkable that the behavior of $\bar{c}$ is equivalent to that obtained from BAM. In this way, we show that the local considerations introduced through MA do not generate a significant impact on the cooperation levels achieved in static systems. However, comparing the results with those of figure $4a$ we notice huge differences in the conditions for cooperation to evolve. In this way we categorically show the importance of the growing process for the emergence of structured cooperative populations.

\bigskip 

\begin{figure}[!hbt] \centering
\includegraphics[angle=0,scale=0.6]{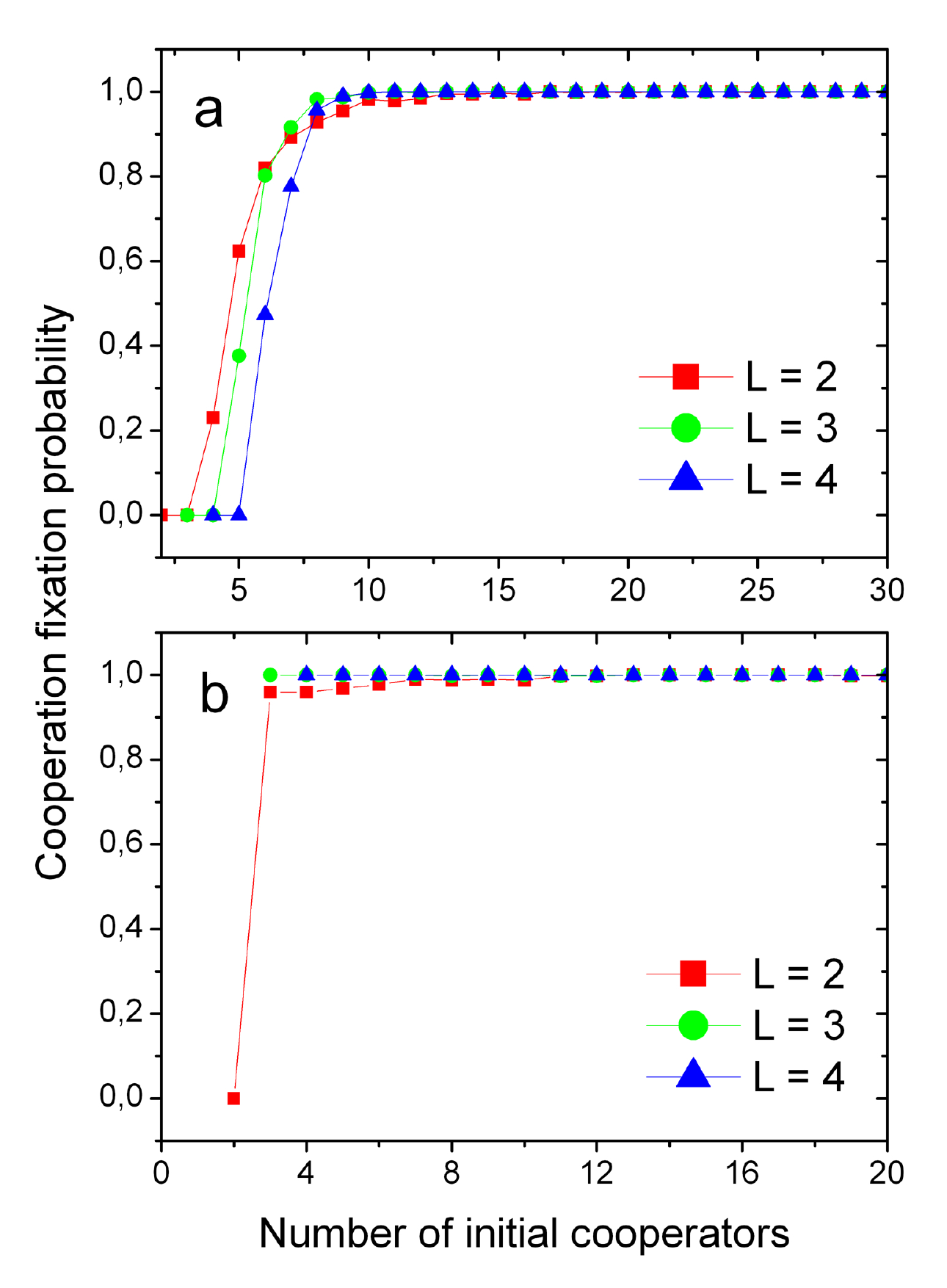}
\caption{The model over networks generated by MA. Cooperation fixation probability as a function of the initial number of cooperators for different values of $L$. We have been considered $r=3,5$ y $M=500$, we choose these values in order to perform a correct comparison with the results showed in figure $2$. a) Without strategies update between the $L$ links introduced between two new individuals. b) Allowing strategies update between the $L$ links.}
\end{figure}

Finally, in Figure $5a$ we show $P_{f}(N_{i})$ for various $L$. Again we observe that $P_{f}$ reaches the value $1$ when the cooperative system reaches a size $N=Nc\sim 20$. Thus we show that MA not only generates a reduction of $r_c$ but also produces a significant decrease in the CS size. This is because there are strategy updates when individuals have less than $L$ connections. However, MA has some limitations in this task when the system is small. In particular, when $N \sim N_0$, MA is equivalent to BAM and therefore there are not strategy updates among the first $L$ connections of the new individual. This makes that defectors have more exploitation capacity when the strategies are updated. To overcome this drawback we not only considered the possibility of strategy updates when the system grows in population but also when it grows in connections. As we defined in Section II, a strategy update is performed each time the system grows a fraction n. Therefore, when the system is large there are several individuals and links introduced to the system before a strategy update. This makes that the results are equivalent if the strategy updates are considered between individuals or between connections. However, when the system is small there are several strategy updates with the incorporation of a single individual. Therefore it becomes important to have the possibility of updates between the $L$ connections introduced to the system for each new individual. In Figure $5c$ we show the cooperation fixation probability $P_{f}(N_{i})$ for different values of $L$ when strategy updates between the $L$ connections are considered. As we shown, this produces a significant reduction of the \emph{Cooperative Seed}. Now $N_c \simeq N_0$ that corresponds to the theoretical minimum that allows MA since the system grows from $N_0$ individuals. Therefore, we have reduced the problem of the emergence of cooperation when it is modeled by the evolutionary prisoner's dilemma to justify the first $N_0$ cooperators. In this way we show the importance of the considerations made in the way in which new individuals and links are introduced to the system.

\section{Potentiality of individuals}

In the evolutionary dynamics employed previously we have considered that all connections have equal weight in the Payoff of individuals. Under this condition, the degree heterogeneities introduce heterogeneities in the Payoffs that favor the emergence of cooperation \cite{N3,N4,N5}. However, a central criticism \cite{N13,N14,N15} of this evolutionary dynamics is that not all connections can be considered equal. In particular, it is often argued that each connection requires the investment of resources such as time, which are finite, and therefore the incorporation of new links does not always increase linearly the potentiality of each individual. We define the potentiality $p_{i}$ of individual $i$ as the maximum theoretical effective Payoff $P_{i}^{eff}$ that an individual can get from its neighborhood divided by $b$. Thus, for the previous results showed we have considered $p_{i}=k_{i}$ assuming that each individual has limitless resources that allow it to increase its potentiality without bound. At the other extreme, we can considered that individuals use all its resources regardless of the number of connections and therefore $p_{i}=1$ for all $k_i$. Therefore, the maximum potentiality is independent of the connectivity of the individual. When this situation is considered \cite{N15} disappear the benefits for the emergence of cooperation resulting from the degree heterogeneity.

\bigskip 

\begin{figure}[!hbt] \centering
\includegraphics[angle=0,scale=0.3]{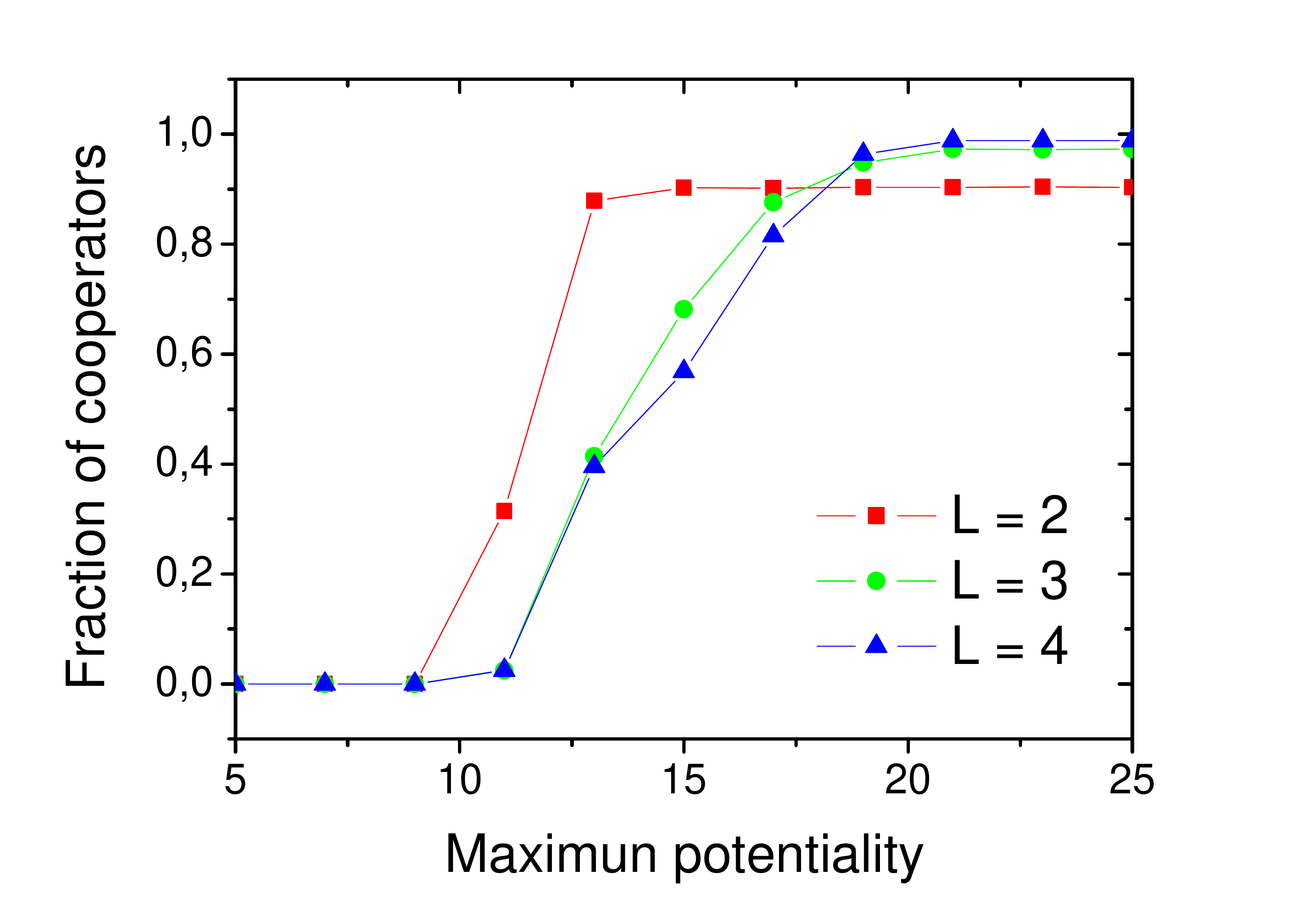}
\caption{Potentiality of individuals. Fraction of cooperators $\bar c$ as a function of the maximum potentiality $K$ for some values of $L$. Results obtained using MA for the network generation. We have been considered $r=2,8$ and averaged the values over $40$ different realizations.}
\end{figure}

\bigskip 

However, it is expected that increasing the connectivity of an individual will enhance its potential in some way. In particular, we can consider the existence of a maximun potentiality $p=K$ determined by those finite resources necessary to interact. We introduced this as follows. On the one hand, if $k_{i}\leq K$ then $p_{i}=k_{i}$ and the effective Payoff $P_{i}^{eff}$ of individual $i$ is equal to $P_{i}$. On the other hand, if $k_{i}>K$ we calculated the effective Payoff as $P_{i}^{eff}=(P_{i}/k_{i})K$. In Figure $6$ we show the fraction of cooperators $\bar{c}$ as a function of $K$ for different values ​​of $L$. It is seen that the system exhibits a phase transition from a non-cooperative to a cooperative behavior for some maximum potentiality $K_c$. It is remarkable that the result is independent of the system size for all $N>N_{c}=N_0$. In addition, we observe that the result does not depend strongly on the mean degree $\bar {k}$. No significant differences have been observed to $L=32$. Figure $6$ shows that when the system growth is considered, it is not necessarily a linear Payoff accumulation with $k$ for cooperation prevails when the system grows. This is remarkable considering the importance of the hubs for the emergence of cooperation when static systems are considered \cite{N3,N4,N5}. Furthermore, when $b$ and $c$ are determined by the total system production and considering that this is independent of the wealth distribution, the consideration of a maximum potentiality produces an increment in $r$ that not just improves the conditions of individuals with $k<K$ but also the conditions for the whole system for cooperation prevail. However, it is still essential that degree heterogeneities increment individual Payoffs in some way. In addition, an important consequence of this result is that it shows the possibility of considering aging and death of individuals without drastically modifying the conditions for cooperation to evolve.

\bigskip 

We can reach similar conclusions by considering the results shown in Figures $2$ and $5$. In them we show that beyond some $N_c$ the system resists the incorporation of defectors when $r>r_{c}$ and, therefore, the system has reached the necessary  heterogeneities for cooperation to prevail. When the system continues to grow from $N_C$, the differences between the Payoff of the new individuals and the more connected one of the system increase. However, this does not reduce the required conditions for cooperation to prevail because, as we showed, $r_{c}\neq r_{c}(N)$. This independence suggests the existence of the critical potentiality $K_{c}$.

\section{Consideration for the first link of individuals}

Although with MA we have introduced greater realism, the system still lacks some topological properties of real systems, such as a high coefficient of degree, which are of great interest. Such structures can be introduced through local considerations in the way the RL are performed, see \cite{N16}. Besides, it is interesting to have some caution in the way that FL are made. Both in BAM and MA, the FL are made by preferential attachment. In a highly cooperative system, with $K$ infinite, the success of individuals is almost proportional to their degree $k$. Therefore, in systems where a greater success in the game leads to more offspring, it is a good approximation to assume that the FL are performed by preferential attachment when new individuals are caused by reproduction of individuals of the system. However, in systems such as the human society a better approximation is to assume the same reproduction rate for all individuals regardless of the success achieved in the game. Moreover, in the case that new individuals arise from the reproduction of individuals in the system, it is a better approximation to assume the FL as random. Therefore, it is of interest to study how this kind of consideration affect to the system. When it is switching from a preferential attachment to a random FL there is a decrease in the degree heterogeneity of the system and, therefore, it is expected that conditions for cooperation to prevail increase. However, it is important to study this situation by comparing the results with those obtained in systems with identical degree distribution but where the FL is performed by preferential attachment. This is important to eliminate undesirable effects arising from differences in the degree distributions. This can be done by considering the following forms of growth. On the one hand, performing the FL random and the other $L-1$ RL with preferential attachment. On the other hand, making the FL with preferential attachment and one of the remaining $L-1$ in a random way. Although a more thorough study is necessary, we simply say that this kind of consideration does not alter the conclusions reached.

\section{Generic cooperative systems}

\begin{figure}[!hbt] \centering
\includegraphics[angle=0,scale=0.37]{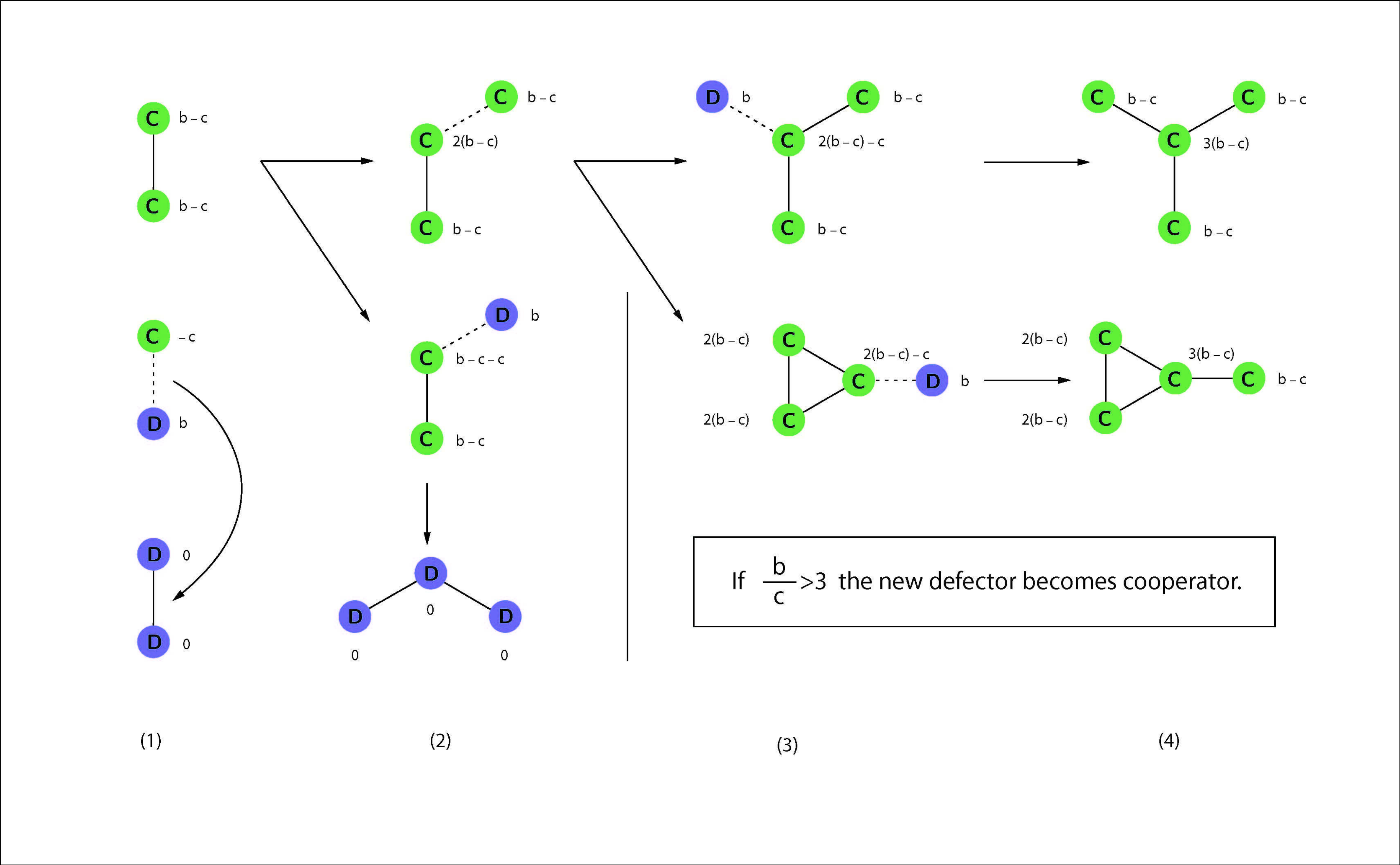}
\caption{Generic cooperative systems. Illustration of the first 4 stages of the system.}
\end{figure}

\bigskip 

In this section we study the model when $n\rightarrow 0$. This condition ensures that the system is in equilibrium when a new individual or connection is introduced to it. In Figure $7$ we illustrate the first four stages of the system. The first stage represents the situation where the system has two individuals, corresponding to the minimum system where the altruistic act can exist. In this case there are three possible combinations of strategies: $C-C$, $C-D$ and $D-D$. For cases $C-C$ and $D-D$ the system is in equilibrium since both individuals have equal Payoff and strategy. However, when each individual has a different strategy, the defector has an advantage over the cooperator, given that $b>-c$ for all $b,c$. Hence, the cooperator becomes defector and drives cooperation to extinction. In consequence, the two individuals should be cooperators to preserve cooperation within the system. In this way we show the difficulty of the cooperation problem when framed into Darwinian Theory using the prisoner's dilemma as a metaphor of the conflict. Since it is not possible to incorporate new connections to the system, the next stage corresponds to the incorporation of a new individual with a single connection. This can be cooperator or defector. In the latter case the complete system is invaded by the defectors because defection always gets higher Payoff than cooperation. This shows that the defection is a better strategy when individuals are connected forming a linear chain. In the case that the new individual is a cooperator the system is in equilibrium. Although the system has an individual with a higher Payoff than other ones, nothing can change since all individuals have the same strategy. In this way we show that under the hypothesis of the model at least the first three individuals must be cooperators for cooperation to evolve.

\bigskip 

In the next stage two things can happen: a new connection can arise between the two disconnected individuals or the system can incorporates a new individual. The first case has no impact on the strategies of the system because all individuals share the same one. However, in the second case appears the possibility to connect the new individual to a cooperator with two cooperative links instead of just one. In this case, when the new individual is a defector and $r>r_{c}=3$, the cooperator has a greater Payoff than the defector and therefore it became cooperator. In this way we show the importance that the degree heterogeneity has for the evolution of cooperation. From this stage, the system remains fully cooperative if each new defector is connected to individuals with at least two connections and $r>r_{c}$. In particular, when the formation of new connections is faster than the incorporation of new individuals $r_c$ can be reduced. If new individuals are attached to ones with at least $k$ connections, then $r_{c}>(k+1)/(k-1)$ ensure $\bar{c}=1$. This mechanism allows to form fully cooperative structures of any topology and $N>3$ without line connections. If besides it is considered the elimination of nodes and links the model allows us the formation of networks with line connections and, therefore, any topology.

\bigskip 

For the conditions imposed, defectors always become cooperators, but never the reverse happens, because the individual who is exploited has enough cooperative support from its neighborhood. Therefore, we find that the system resists the invasion of defectors locally. However, as shown throughout this work it can happen that the system has $\bar{c}<1$ and keeps the levels of cooperation when the system grows by new defectors. In this case, the defection can invade part of the system but it has mechanisms that allow it to restore cooperation in the parts affected by the defection. This suggests that the mechanisms by which the system resists the defection invasions can range from local, where there are involved the exploited individual and its neighborhood, to global, where the whole system acts against defection as it happens when it is considered static systems with $P_c=1/2$.

\section{The first three cooperators}

In the previous section we reduced the \emph{Cooperative Seed} to three individuals. In this, we perform biological considerations that allow us to justify this structure. For this we consider that individuals have the potential of reproduction. This feature is ubiquitous in biological individuals except in some cases, as worker ants. If we also consider that reproduction is sexual, we can justify the cooperation of the first two individuals. If there is not cooperation for reproduction there are not offspring and the extinction of the species is assured. Therefore, in species with sexual reproduction just evolve those individuals who are cooperators at least to reproduce. In this case, the new individuals are a consequence of the genetic reproduction. In the simplest case where the two cooperative individuals only have one child with a single initial link, the problem is illustrated in the second stage of Figure $7$ where cooperation is extinguished. As we showed throughout this work, cooperation can emerge when there is heterogeneity between the new individual and the rest of the system. In this case we can introduce the heterogeneity considering that offspring has less intrinsic adaptability ($IA$) to the environment than their parents, where $IA$ is the adaptability of the individual living alone. Therefore, if we consider that adults have $AI = 0$ then the son has $AI=-d(t)$, where $d>0$. It is expected that $IA$ grows in time reaching $d=0$ when the offspring become adult. 

\bigskip 

\begin{figure}[!hbt] \centering
\includegraphics[angle=0,scale=0.4]{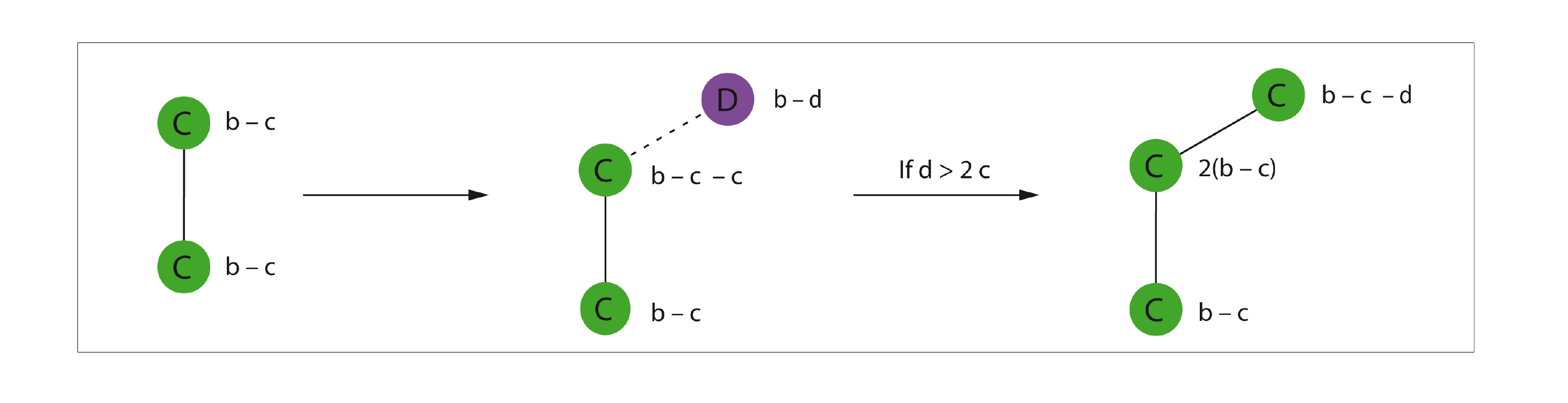}
\caption{The first three cooperators. The son becomes cooperator.}
\end{figure}

\bigskip 

In Figure $8$ we schematically place the mechanism by which the offspring becomes cooperator. We consider the case where there is only one son with only a link to one parent. In the first stage we show the two individuals that reproduce and, therefore, initially cooperators. The system reaches the second stage when the son is conceived. It is important to note that this analysis can be extended to the case of several sons as well as many links for each new individual. As we shown in Figure if $d>2c$ the cooperator has a higher Payoff than the new individual and therefore cooperation is a better strategy. This occurs until the offspring grows to reach the condition $d=2c$, from which defection is better. In our case $n\rightarrow 0$ and thus it is assured that the new individual becomes cooperators imitating the strategy of their parents. In this way we show conditions to form the first three cooperators.

\section{Conclusions}

In this paper we study the cooperation problem framed within the evolutionary game theory using the prisoner's dilemma as a metaphor of the problem. Considering the growing process of the system and individuals with imitation capacity, we show conditions that allow forming highly cooperative systems of any size and topology when the possibility of spontaneous changes of strategy are neglected. Performing considerations ​​on the formation of links and the incorporation of individuals, we reduce the required benefit-cost ratio for cooperation to evolve approaching to the theoretical minimum $r=1$, when the mean connectivity of the individuals is increased. Also, we distinguish different mechanisms that allow the system to maintain high levels of cooperation when the system grows by incorporation of defectors. These mechanisms require heterogeneity among individuals for cooperation to evolve. However, the required conditions and heterogeneities are drastically reduced as compared to those required for static networks. In this way, we show that the growth process of the system is essential for the evolution of cooperation when individual have imitation capacity. These results together with the those showed in \cite{N15} give an answer to the first question of the new formulation of the cooperation problem carried out in the introduction. Therefore we shed light to the understanding of the emergence and persistence of cooperation in structured populations within the Darwinian Theory. In particular, this work could help to understand the evolution of our society as well as its structure. Besides, it could be important for the development of a more egalitarian economy that improves the quality of life of our society. Finally, the results have important consequences in order to take into account the possibility of mutations, since allow it study the problem assuming a fully cooperative system as initial condition.

\section{Acknowledgements}

I acknowledge financial support from Generalitat de Catalunya under grant $2009SGR00164$ and PhD fellowship $2011FIB787$. I am grateful to the Statistical Physics Group of the Autonomous University of Barcelona. In particular, I would like to thank David Jou and Vicen\c c M\'{e}ndez for their unconditional support and assistance with the manuscript. To Carolina Perez Mora for complete my life. To Lucas Gomez Portillo for showing me the importance of never giving up in adversity.

\end{document}